\documentstyle[prl,aps,preprint,tighten,floats]{revtex}
\begin{document}
\draft
\preprint{UPR-672-T}
\date{July 1995}
\title{Dyonic BPS Saturated Black Holes of Heterotic String on
a Six-Torus} \author{Mirjam Cveti\v c
\thanks{E-mail address: cvetic@cvetic.hep.upenn.edu}
and Donam Youm\thanks{E-mail address: youm@cvetic.hep.upenn.edu}}
\address {Physics Department \\
          University of Pennsylvania, Philadelphia PA 19104-6396}
\maketitle
\begin{abstract}
{Within effective heterotic superstring theory compactified on a
six-torus we derive minimum energy (supersymmetric), static,
spherically symmetric solutions, which are manifestly invariant
under the target space $O(6,22)$ and the strong-weak coupling
$SL(2)$ duality symmetries with 28 electric and 28 magnetic
charges subject to one constraint. The class of solutions with a
constant axion corresponds to dyonic configurations subject to two charge
constraints, with purely electric [or purely magnetic] and dyonic
configurations preserving ${1\over 2}$ and  ${1\over 4}$
of $N=4$ supersymmetry, respectively.  General dyonic configurations
in this class have a space-time of extreme Reissner-Nordstr\" om black
holes while configurations with more constrained charges have
a null or a naked singularity.}
\end{abstract}

There has been an accumulating evidence for the strong-weak coupling
duality (referred to as the $S$ duality) in string theory
(See for example  Refs. \cite{WITTENI,HTI}), which relates
supersymmetric vacua of a strongly coupled theory to supersymmetric
vacua of a dual $-$ weakly coupled $-$ theory.  Deeper understanding
of these duality symmetries would provide us with a handle on the
non-perturbative nature of superstring theory.

In four-dimensions, string vacua with $N=4$ low energy supersymmetry
are conjectured to be self-dual, {\it i.e.}, the string vacua
of the heterotic string compactified on a six-torus transform into each
other under the $SL(2,Z)$ transformations
\footnote{The heterotic string compactified on a six-torus is
conjectured to be dual to the Type IIA string compactified on a
$T^2 \times K_3$ surface; this duality has its origin in the
string-string duality  conjecture \cite{WITTENI,HTI,DUFFSS,STRDUAL}
of the heterotic and the Type IIA string theory in six-dimensions.}.
The $SL(2,Z)$ symmetry acts on charges, the axion and the
four-dimensional dilaton field, whose value determines the string
loop expansion parameter and parameterizes the strength of the string
coupling.  In addition, the string world-sheet action can
be cast in a manifestly $O(6,22)$ symmetric form \cite{TARGET,SCHWARZ},
referred to as the target space $T$ duality symmetry.

Evidence for the $S$ duality conjecture of $N=4$ supersymmetric string vacua
has been provided by demonstrating the $S$ duality invariance of quantities
which are believed not to be modified by string quantum corrections,
{\it e.g.}, the low energy effective field theory \cite{DEROO,SCHWARZ},
allowed spectrum of electric and magnetic charges, Yukawa couplings
between massless scalars and massive charged states as well as the
Bogomol'nyi-Prasad-Sommerfeld (BPS) saturated mass spectrum of the
corresponding non-trivial configurations in the effective theory
\cite{HARVEY,SCHSEN}.

The BPS saturated states within the effective theory compactified on
a six-torus have been addressed for states with special charge
configurations (See for example Refs.
\cite{HARVEY,KALLOSH,BANKS,HMON,SEN1}), which in turn prevented
one from establishing the full symmetry structure of such
configurations as well as the full nature of their singularity
structure.  In this paper, we shall present the explicit form of general
BPS saturated (supersymmetric), spherically symmetric, static
configurations in the effective heterotic string theory compactified on a
six-torus at generic points of moduli space, which can be obtained from
the generating solution with, among scalar fields, only diagonal internal
metric and the dilaton turned on.  The BPS saturated spectrum is both
$O(6,22)$ and $SL(2,I\!\!R)$ invariant
\footnote{At the quantum level, these symmetries are integer valued.}.
In addition, the explicit form of these configurations allows for a
synthetic analysis of their singularity structures and their
thermal properties within the class of solutions.

The effective field theory of massless bosonic fields for heterotic
string on a Narain torus \cite{NARAIN} can be obtained by compactifying the
ten-dimensional $N=1$ supergravity theory coupled to $N=1$ super-Maxwell
theory on a six-torus \cite{SCHWARZ,SEN2}.
The ten-dimensional bosonic fields are given by $\hat{G}_{MN}$,
$\hat{B}_{MN}$, $\hat{A}^I_M$ and $\Phi$ ($0 \le M,N \le 9$,
$1 \le I \le 16$), which correspond to ten-dimensional metric,
two-form field, gauge fields of $U(1)^{16}$, and the dilaton field,
respectively.  The field strengths of $\hat{A}^I_M$ and $\hat{B}_{MN}$ are
defined as $\hat{F}^I_{MN} = \partial_M \hat{A}^I_N - \partial_N
\hat{A}^I_M$ and $\hat{H}_{MNP} = \partial_M \hat{B}_{NP} -
{1\over 2}\hat{A}^I_M \hat{F}^I_{NP} + {\rm cyc.\ perms.}$, respectively.

The Kaluza-Klein compactification \cite{KAL} of the original
ten-dimensional action on a six-torus is obtained with the following
Ansatz for the Zehnbein:
$\hat{E}^A_M  = \left ( \matrix{e^{\phi \over 2}e^{\alpha}_{\mu} &
A^{(1)\,m}_{\mu}e^a_m \cr 0 & e^a_m} \right )$,
where $A^{(1)\,m}_{\mu}$ ($m=1,...,6$) are Kaluza-Klein $U(1)$ gauge fields
and $\phi \equiv \Phi - {\rm ln}\,{\rm det}\, e^a_m$ is the four-dimensional
dilaton field.

The four-dimensional action \cite{SCHWARZ,SEN2} for massless bosonic fields
contains the following fields:
the graviton $g_{\mu\nu}$, the dilaton $\phi$, 28 $U(1)$ gauge fields
${\cal A}^i_{\mu} \equiv (A^{(1)\, m}_{\mu}, A^{(2)}_{\mu\, m},
A^{(3)\, I}_{\mu})$ defined as $A^{(2)}_{\mu\,m} \equiv \hat{B}_{\mu\,m}
+ \hat{B}_{mn}A^{(1)\,n}_{\mu} + {1\over 2}a^I_m A^{(3)\,I}_{\mu}$,
$A^{(3)\,I}_{\mu} \equiv \hat{A}^I_{\mu} - a^I_m A^{(1)\,m}_{\mu}$ with the
field strengths ${\cal F}^i_{\mu\nu} = \partial_{\mu} {\cal A}^i_{\nu}
- \partial_{\nu} {\cal A}^i_{\mu}$, the two form field
\footnote{The four-dimensional two-form field is equivalent to a
pseudo-scalar (the axion)  $\Psi$ through the duality transformation
$H^{\mu\nu\rho} = -{{e^{2\phi}} \over {\sqrt{-g}}}
\varepsilon^{\mu\nu\rho\sigma}\partial_{\sigma}\Psi$.}
$B_{\mu\nu}$ with the field strength given by
$H_{\mu\nu\rho}= e^{\alpha}_{\mu} e^{\beta}_{\nu} e^{\gamma}_{\rho}
\hat{E}^M_{\alpha} \hat{E}^N_{\beta} E^P_{\gamma} \hat{H}_{MNP} =
\partial_{\mu} B_{\nu\rho} - {1\over 2} {\cal A}^i_{\mu}L_{ij}
{\cal F}^j_{\nu\rho} + {\rm cyc. perms.}$, and a symmetric $O(6,22)$
matrix $M$ of scalar fields, which can be expressed in terms of
the following $O(6,22)$ matrix
\begin{equation}
V = \left ( \matrix{V^I \cr V^{II} \cr V^{III}} \right ) =
\left ( \matrix{E^{-1} & -E^{-1}C & -E^{-1}a^T \cr
0 & E & 0 \cr 0 & a & I_{16}} \right )
\label{mviel}
\end{equation}
as $M = V^T V$, where $E \equiv [e^a_m]$, $C \equiv [{1\over 2}
\hat{A}^I_m \hat{A}^I_n + \hat{B}_{mn}]$ and $a \equiv [\hat{A}^I_m]$.
$V$ plays a role of a Vielbein in the $O(6,22)$ target space.

The 4-d effective action is invariant under the $O(6,22)$
transformations \cite{SCHWARZ,SEN2}:
\begin{equation}
M \to \Omega M \Omega^T ,\ \ \ {\cal A}^i_{\mu} \to \Omega_{ij}
{\cal A}^j_{\mu}, \ \ \ g_{\mu\nu} \to g_{\mu\nu}, \ \ \ \phi \to \phi .
\label{tdual}
\end{equation}
Here, $\Omega \in O(6,22)$, {\it i.e.}, $\Omega^T L \Omega = L$,
where $L$ is an $O(6,22)$ invariant matrix.  In addition, the
corresponding equations of motion and Bianchi identities have the
invariance under the $SL(2,I\!\!R)$ transformations \cite{STRWK,SEN2}:
\begin{equation}
S \to S^{\prime}={{aS+b}\over{cS+d}},\ M\to M ,\ g_{\mu\nu}\to g_{\mu\nu},\
{\cal F}^i_{\mu\nu} \to {\cal F}^{\prime\, i}_{\mu\nu} =
(c\Psi + d){\cal F}^i_{\mu\nu} + ce^{-\phi} (ML)_{ij}
\tilde{\cal F}^j_{\mu\nu},
\label{sdual}
\end{equation}
where $S \equiv \Psi + i e^{-\phi}$, $\tilde{\cal F}^{i\,\mu\nu} =
{1\over 2}(\sqrt{-g})^{-1} \varepsilon^{\mu\nu\rho\sigma}
{\cal F}^i_{\rho\sigma}$ and $a,b,c,d \in I\!\!R$ satisfy $ad-bc=1$.

The ten-dimensional supersymmetry transformations for
gravitino $\psi_M$, dilatino $\lambda$ and 16 gaugini $\chi^I$ are
expressed in terms of the four-dimensional fields as:
\begin{eqnarray}
\delta \hat{\psi}_{\mu} &=&
\partial_{\mu} \varepsilon + {1\over 4}\omega_{\mu\beta\gamma}
\gamma^{\beta\gamma}\varepsilon + {1\over 4}e^{\alpha}_{\mu}
\eta_{\alpha[\beta}e^{\nu}_{\gamma]}\partial_{\nu} \phi
\gamma^{\beta\gamma}\varepsilon
- {1\over 8}e^{-\phi}H_{\mu\nu\rho}\gamma^{\nu\rho}\varepsilon
\nonumber\\ &+& {1\over 8}(e^n_b \partial_{\mu}
e_{nc} - e^n_c\partial_{\mu}e_{nb})I\otimes \Gamma^{bc}\varepsilon
- {1\over 4}e^{-{\phi \over 2}}[(VL)^{I}_{c\,i}
+(VL)^{II}_{c\,i}]{\cal F}^i_{\mu\nu}\gamma^{\nu 5}\otimes
\Gamma^c \varepsilon ,
\nonumber\\
\delta \psi_d &=& -{1\over 4}e^{-{\phi \over 2}}[e^m_d\partial_{\mu}
e_{mb} + e^m_b \partial_{\mu} e_{md} - e^m_d e^n_b(\partial_{\mu}B_{mn}
+ {1\over 2}a^I_m \partial_{\mu}a^I_{n} - {1\over 2}a^I_n
\partial_{\mu}a^I_m)] \gamma^{\mu 5} \otimes \Gamma^b \varepsilon \nonumber\\
&-&{1\over 8}e^{-\phi}[(VL)^I_{d\,i} - (VL)^{II}_{d\,i}]{\cal F}^i_{\mu\nu}
\gamma^{\mu\nu}\varepsilon ,
\nonumber\\
\delta \lambda &=& e^{-{\phi \over 2}}\partial_{\mu} \Phi \gamma^{\mu}
\varepsilon - {1\over 6} e^{-{3\over 2}\phi}H_{\mu\nu\rho}
\gamma^{\mu\nu\rho}\varepsilon
- {1\over 2}e^{-{\phi \over 2}}(\partial_{\mu}B_{mn} +
{1\over 2}a^I_m \partial_{\mu} a^I_n-{1\over 2}a^I_n \partial_{\mu}
a^I_m)\gamma^{\mu} \otimes \Gamma^{mn} \varepsilon  \nonumber\\
&-& {1\over 2}e^{-\phi}(VL)^I_{d\,i} {\cal F}^i_{\mu\nu}
\gamma^{\mu\nu 5} \otimes \Gamma^d \varepsilon ,
\nonumber\\
\delta \chi^I &=& 2e^{-{\phi\over 2}}\partial_{\mu} a^I_m \gamma^{\mu 5}
\otimes \Gamma^m \varepsilon + e^{-\phi}(VL)^{III}_{I\,i} {\cal F}^i_{\mu\nu}
\gamma^{\mu\nu} \varepsilon ,
\label{foursusy}
\end{eqnarray}
where $\delta \hat{\psi}_{\mu} \equiv \delta \psi_{\mu}-A^{(1)\,m}_{\mu}
\delta \psi_m$ and $\delta \psi_d \equiv e^m_d \delta \psi_m$.
Here, $\gamma^{\alpha}$ and $\Gamma^{a}$ satisfy the $O(1,3)$ and $O(6)$
Clifford algebras, respectively.  The gamma matrices
with curved indices are defined as $\gamma^{\mu}\equiv
e^{\mu}_{\alpha}\gamma^{\alpha}$ and $\Gamma^m \equiv e^m_a \Gamma^a$.

Static configurations, which saturate the Bogomol'nyi bound for their
masses, {\it i.e.}, the minimum energy configurations in their class,
correspond to bosonic backgrounds which preserve the supersymmetry
transformations (\ref{foursusy}) \cite{HARVEY}, thus also referred to
as supersymmetric configurations.  The Killing spinor equations, which
are obtained by setting the supersymmetry transformations
(\ref{foursusy}) to zero, provide constraints on the Killing spinors
$\varepsilon$ and a set of coupled first order differential equations
for the supersymmetric bosonic backgrounds.

Our aim is to obtain general supersymmetric, spherically symmetric,
static configurations with a general allowed charge content associated with
the 28 $U(1)$ gauge fields.  The four-dimensional space-time metric
is chosen to be of the form:
\begin{equation}
g_{\mu\nu} dx^{\mu} dx^{\nu} = \lambda dt^2 -
\lambda^{-1}dr^2 - R(d\theta^2 + {\rm sin}^2 \theta d\phi^2),
\label{sphmet}
\end{equation}
and the scalar fields $M$, $\phi$ and $\Psi$ depend on the radial
coordinate $r$, only.  The Maxwell's equations and Bianchi
identities determine the $U(1)$ field strengths to be
\begin{equation}
{\cal F}^i_{tr} = {e^{\phi} \over R}[M_{ij}\tilde{Q}_j + \Psi(ML)_{ij}
P_j], \ \ \ {\cal F}^i_{\theta\phi} = P_i\, {\rm sin}\,\theta ,
\label{elecmag}
\end{equation}
where $P_i$'s correspond to the physical magnetic charges and the physical
electric charges \cite{WITTENII} are given by $Q_i = e^{\phi_{\infty}}
[M_{ij\,\infty}\tilde{Q}_j + \Psi_{\infty}(ML)_{ij\,\infty}P_j]$.

One can show that with the above static, spherically symmetric Ansatz the
Killing spinors are invariant under the $O(6,22)$ transformations and
transform as $\varepsilon \rightarrow [{\rm cos} (\Delta/2) + i\gamma^5
{\rm sin} (\Delta /2)]\varepsilon$ under the $SL(2,I\!\!R)$.
Here, ${\rm tan} \Delta = -ce^{-\phi}/(c\Psi+d)$.
The first order differential equations are thus invariant
under both transformations and therefore one can generate new class of
supersymmetric solutions by imposing $O(6,22)$ and $SL(2,I\!\!R)$
transformations on the known supersymmetric solution.  One can bring
the arbitrary asymptotic values of the scalar fields to the forms
$M_{\infty}=I$ and $S_{\infty}=i$ by imposing the following $O(6,22)$
and $SL(2,I\!\!R)$ transformations:
\begin{equation}
M_{\infty} \to \hat{M}_{\infty}=\hat{\Omega}M_{\infty}\hat{\Omega}^T=I,
\ \ \ \ \ \  S_{\infty} \to \breve{S}_{\infty} = (aS_{\infty}+b)/d = i,
\label{asympt}
\end{equation}
where $\hat{\Omega} \in O(6,22)$, $ad=1$, and in the quantized theory
the charge lattice vectors live in the new transformed lattice.
Then, the subsets of $O(6,22)$ and $SL(2,I\!\!R)$ transformations
that preserve the above new asymptotic values of $M$ and $S$ are
$O(6)\times O(22)$ and $SO(2)$, respectively.  To obtain solutions
with arbitrary asymptotic values of $M$ and $S$, one has to undo
the above transformations.

We are going to find the general solution for configurations where,
from the scalar fields, only the diagonal internal metric and the
dilaton field are non-zero.  We shall refer to such configurations
as generating ones, since all the other configurations in this class
can be obtained by performing a subset of $O(6)\times O(22) \subset
O(6,22)$ and $SO(2) \subset SL(2,I\!\!R)$ transformations on the
generating ones.  Note, that configurations obtained in that manner
have the same four-dimensional space-time structure and thus the same
singularity and thermal properties as the generating solution.

The Killing spinor equations for the configuration with only non-zero
scalar fields given by the diagonal internal metric
($e^m_a = \delta^m_a e_a$) and the dilaton take the form:
\begin{eqnarray}
\sqrt{\lambda}R[\partial_r {\rm ln}\,\lambda + \partial_r \phi]
\varepsilon_{u,\ell}
&=& \pm\Sigma_{a=1}^6({\bf Q}^{(1)}_a + {\bf Q}^{(2)}_a)\Gamma^a
\varepsilon_{\ell,u},
\nonumber\\
\sqrt{\lambda}R[\partial_r {\rm ln}\,\lambda -
\partial_r \phi]\varepsilon_{u,\ell} &=& i\Sigma_{a=1}^6({\bf P}^{(1)}_a
+ {\bf P}^{(2)}_a) \Gamma^a \varepsilon_{\ell,u},
\nonumber\\
2\sqrt{\lambda}R\partial_r {\rm ln}\,e_a \Gamma^a \varepsilon_{u,\ell} &=&
[{\mp}({\bf Q}^{(1)}_a - {\bf Q}^{(2)}_a)
+i({\bf P}^{(1)}_a - {\bf P}^{(2)}_a)]\varepsilon_{\ell,u},\ \ \
a=1,\cdots ,6 ,
\nonumber\\
\partial_r\sqrt{\lambda\,R}&=&0,
\label{killing}
\end{eqnarray}
where
${\bf Q}^{(1)}_a \equiv e^{\phi \over 2} e^m_a \tilde{Q}_m$,
${\bf Q}^{(2)}_a \equiv e^{\phi \over 2} e^a_m \tilde{Q}_{6+m}$,
${\bf P}^{(1)}_a \equiv e^{-{\phi \over 2}} e^a_m P_m$, and
${\bf P}^{(2)}_a \equiv e^{-{\phi \over 2}} e^m_a P_{6+m}$.
And from $\delta \chi^I = 0$, one has $P^{(3)}_I = 0 = Q^{(3)}_I$.
It can be shown \cite{SUPER} that out of $2\cdot 28$ dyonic charges,
only two magnetic and two electric charges can be non-zero with electric
and magnetic charges arising from different $U(1)$ factors, with one
set of electric and  magnetic charges arising from the Kaluza-Klein sector
and the other set arising from the two-form gauge fields with
the same corresponding indices.  Without loss of generality, we choose
the non-zero charges to be $P^{(1)}_1, P^{(2)}_1, Q^{(1)}_2, Q^{(2)}_2$.

The upper $\varepsilon_{u}$ and lower $\varepsilon_{\ell}$
two-component spinors are subject to the constraints:
$\Gamma^1 \varepsilon_{u,\ell}=i\eta_P\varepsilon_{\ell,u}$
if ${\bf P}^{(1)}_1 \ne 0 $ and/or ${\bf P}^{(2)}_1 \ne 0$, and
$\Gamma^2\varepsilon_{u,\ell}=\mp \eta_Q \varepsilon_{\ell,u}$ if
${\bf Q}^{(1)}_2 \ne 0 $ and/or ${\bf Q}^{(2)}_2 \ne 0$.  Here, $\eta_P$ and
$\eta_Q$ are  $\pm 1$.  Note, that non-zero magnetic and electric charges
each  break ${1\over 2}$ of the remaining supersymmetries.  Thus, purely
electric [or magnetic] configurations preserve ${1\over 2}$, while
dyonic solutions preserve ${1\over 4}$ of $N=4$ supersymmetry in
4 dimensions.  The first and the second sets of configurations fall into
vector- and hyper-supermultiplets, respectively.

The explicit form for the static, spherically symmetric generating
solution is given by
\footnote{Such a solution is obtained along the similar
lines as the generating solution for the supersymmetric, spherically
symmetric solutions in Abelian Kaluza-Klein theory \cite{SUPER}.}:
\begin{eqnarray}
\lambda &=& r^2/[(r-\eta_P P^{(1)}_{1})(r-\eta_P P^{(2)}_{1})
(r- \eta_Q Q^{(1)}_{2})(r-\eta_Q Q^{(2)}_{2})]^{1\over 2},
\nonumber\\
R &=& [(r-\eta_P P^{(1)}_{1})(r - \eta_P P^{(2)}_{1})
(r - \eta_Q Q^{(1)}_{2})(r-\eta_Q Q^{(2)}_{2})]^{1\over 2},
\nonumber\\
e^{\phi}&=&\left [{(r-\eta_P P^{(1)}_{1})
(r- \eta_P P^{(2)}_{1})} \over {(r- \eta_Q Q^{(1)}_{2})
(r- \eta_Q Q^{(2)}_{2})}\right]^{1\over 2},
\nonumber\\
g_{11}&=&\left ({{r- \eta_P P^{(2)}_{1}} \over {r-\eta_P P^{(1)}_{1}}}
\right ), \
g_{22}=\left ({{r- \eta_Q Q^{(1)}_{2}} \over {r- \eta_Q Q^{(2)}_{2}}}
\right ),\
g_{mm}=1\ \ \ (m \neq 1,2).
\label{gensol}
\end{eqnarray}
Here, the radial coordinate is chosen so that the horizon is at $r=0$.
The requirement that the ADM mass of the above configuration saturates
the Bogomol'nyi bound restricts the choice of parameters $\eta_{P,Q}$
such that $\eta_P {\rm sign}(P^{(1)}_1+P^{(2)}_1)=-1$ and $\eta_Q
{\rm sign}(Q^{(1)}_2+Q^{(2)}_2)=-1$, thus yielding non-negative BPS
saturated ADM mass of the form
$M_{BPS} = |P^{(1)}_{1}+ P^{(2)}_{1}|+|Q^{(1)}_{2}+ Q^{(2)}_{2}|$.
In order to have regular BH solution with singularity behind or on
the horizon, one has to choose the relative signs of two magnetic
and two electric charges to be the same
\footnote{Note, that the case of opposite relative signs for the two
electric charges [and two magnetic charges] and the equal magnitude
of the two electric [and two magnetic charges] would yield
zero ADM mass as pointed out in a related context by Hull and Townsend
\cite{HTII}. Such purely electrically (or purely magnetically)
charged configurations were found and studied in Refs. \cite{Behrndt}, while
dyonic ones and their implications for enhanced symmetries\cite{HTII}
at special points of moduli space were addressed in Ref.\cite{CYS}.
Such configurations are {\it not} regular; they have a naked singularity.}.
Thus, the solution has {\it always} nonzero BPS saturated ADM mass.

A general class of solutions with zero axion can be obtained from the
generating ones by performing a subset of $O(6)\times O(22) \subset
O(6,22)$ transformations which generate new types of solutions from
the generating one.  These transformations correspond to $SO(6)/SO(4)$
transformations with ${{6\cdot 5 - 4\cdot 3}\over 2}=9$
parameters and $SO(22)/SO(20)$ transformations with ${{22\cdot 21 -
20\cdot 19} \over 2} = 41$ parameters, which along with the original 4
charges provide a configuration with $56-2=54$ charges;
namely, those are configurations with 28 electric ${\vec Q}$ and
28 magnetic ${\vec P}$ charges subject to the following two
constraints (in the basis where the asymptotic value of $M$ takes
an arbitrary value):
\begin{equation}
{\vec P}^T{\cal M}_{\pm}{\vec Q}=0\ \ \ \
({\cal M}_{\pm} \equiv (LML)_{\infty}\pm L).
\label{gencon}
\end{equation}

The $SO(2) \subset SL(2,I\!\!R)$ transformation provides one with one
more parameter ${\rm tan}\Delta = -ce^{-\phi}/(c\Psi+d)$, which
replaces the two constraints (\ref{gencon}) with one $SL(2,I\!\!R)$ and
$O(6,22)$ invariant constraint on charges
:
\begin{equation}
{\vec P}^T{\cal M}_{-}{\vec Q}\,[{\vec Q}^T{\cal M}_{+}{\vec Q} -
{\vec P}^T{\cal M}_{+}{\vec P}] - (+ \leftrightarrow -) = 0.
\label{fgencon}
\end{equation}
Therefore, the general configurations in this class have $2\cdot 28 - 1 = 55$
charge degrees of freedom
\footnote{The constraint (\ref{fgencon}) on charges signals that the
obtained class of configurations may not be the most general
supersymmetric one. This constraint is removed  for the supersymmetric
[non-extreme] states by  applying an additional subset of $SO(8,24)$
transformations \cite{SEN1,SEN5} on the corresponding  supersymmetric
[non-extreme] generating solutions.  $O(8,24)$ is the symmetry of the
effective three-dimensional action for the corresponding stationary
solutions.  Analogous procedure was used \cite{CYALL} to generate all
the black holes in Abelian Kaluza-Klein theory.}.

The ADM mass for a general configuration in this class can be
obtained from the one for the generating solutions and can be cast in
the following $O(6,22)$ and $SL(2,I\!\!R)$ invariant form
\footnote{We thank A. Sen for pointing out to us the procedure
to derive such a mass.}:
\begin{equation}
M^2_{BPS} = e^{-\phi_\infty} \left\{{\vec P}^T {\cal M}_+ {\vec P} +
{\vec Q}^T {\cal M}_+ {\vec Q}+ 2\left [({\vec P}^T{\cal M}_+{\vec P})
({\vec Q}^T{\cal M}_+{\vec Q})-({\vec P}^T{\cal M}_+
{\vec Q})^2\right]^{1\over 2}\right\}.
\label{Bogmass}
\end{equation}
Note, that when the magnetic $\vec{P}$ and electric $\vec{Q}$ charges
are parallel in the $SO(6,22)$ sense, this ADM mass is the bound
for configurations that preserve ${1\over 2}$ of $N=4$ supersymmetry
\cite{HARVEY,BANKS,HMON,SEN1,GP,PEET},
{\it i.e.}, the corresponding generating solution is either purely
electric or purely magnetic.  In the case when the magnetic and
electric charges are not parallel, the mass is larger and the
configurations preserve $1\over 4$ of $N=4$ supersymmetry.

We now turn to the discussion of the thermal and global space-time
properties of such configurations, which can be classified according
to the number of non-zero charges of the generating
solutions:
\begin{itemize}
\item
The case with {\it all the four charges non-zero}
\footnote{Solutions with two electric [and two magnetic]
charges equal correspond to configurations with constant $M$.
A class of such configurations was obtained by Kallosh {\it et al.}
\cite{KALLOSH}.  For the case where all the four charges are equal, all
the scalars are constant and the four-dimensional metric reduces to that of
Reissner-Nordstr\" om BH's, which is also pointed out in Ref.\cite{DUFFR}.}
corresponds to BH's with a horizon at $r=0$ and a time-like
singularity hidden behind the horizon, {\it i.e.}, the global space-time
structure is that of the extreme Reissner-Nordstr\" om BH's.
The corresponding Hawking temperature $T_H = \partial_r
\lambda |_{r=0}/(2\pi)$ is zero and the entropy ( $\equiv$
${1\over 4}$ of the area of the event horizon) is finite
$S=\pi\sqrt{|P^{(1)}_{1} P^{(2)}_{1} Q^{(1)}_{2} Q^{(2)}_{2}|}$.
\item
The case with three nonzero charges corresponds to solutions with a
singularity located at the horizon ($r=0$), $T_H=0$ and $S=0$.
\item
The case with two nonzero charges
\footnote{Configurations with two non-zero electric charges were
constructed by Sen \cite{SEN1}, and shown by Peet \cite{PEET} to be
supersymmetric.  The supersymmetric configurations with
$P^{(1)}_{1}\ne 0$ and $Q^{(1)}_{2}\ne 0$,
found by the authors \cite{SUPER}, correspond to configurations in the
Kaluza-Klein sector of the theory.},
say, $P^{(1)}_{1} \ne 0 \ne P^{(2)}_{1}$,
corresponds to singular solutions with the horizon and the singularity
coinciding at $r=0$, $T_H=1/(4\pi \sqrt{|P^{(1)}_{1} P^{(2)}_{1}|})$
and $S=0$.
\item
The case with one nonzero charge
\footnote{$P^{(1)}_{1} \ne 0$ case corresponds to the
Kaluza-Klein monopole solution of Gross and Perry, and Sorkin \cite{GPS},
and were shown to be supersymmetric by Gibbons and Perry\cite{GP}.
The case when $P^{(2)}_{1} \ne 0$ corresponds to the
$H$-monopole solution \cite{HMON}.}
corresponds to BH's with a naked singularity, $T_H=\infty$ and $S=0$.
\end{itemize}

\acknowledgments
The work is supported by U.S. DOE Grant No. DOE-EY-76-02-3071, the
NATO collaborative research grant CGR 940870 and the National Science
Foundation Career Advancement Award PHY95-12732. M.C. would like to
thank M. Duff, G. Gibbons, J. Harvey, C. Hull, D. L\" ust and especially
A. Sen for useful discussions and the Aspen Center for Theoretical
Physics for hospitality during the completion of the work.

\end{document}